\documentclass[copyright,creativecommons]{eptcs}
\providecommand{\event}{THedu'11} % Name of the event you are submitting to
\usepackage{breakurl}             % Not needed if you use pdflatex only.

\usepackage[T1]{fontenc}
\usepackage[utf8]{inputenc} 
\usepackage{graphicx}
\usepackage{mathbbol}
\usepackage{amsfonts}
\usepackage{alltt}
\usepackage{color}
\usepackage{epic}
\usepackage{eepic}
\usepackage{csquotes}

\title{Computer-Assisted Program Reasoning Based on a \\ 
Relational Semantics of Programs\thanks{This research was funded by the 
Austrian Science Fund (FWF): W1214-N15, project DK10.}}
\author{Wolfgang Schreiner
\institute{Research Institute for Symbolic Computation (RISC)\\
Johannes Kepler University\\
Linz, Austria}
\email{Wolfgang.Schreiner@risc.jku.at}
}
\def\titlerunning{Computer-Assisted Program Reasoning Based on a Relational Semantics of Programs}
\def\authorrunning{Wolfgang Schreiner}

\newcommand{\ffont}[1]{\textsf{#1}}
\newcommand{\pfont}[1]{\texttt{#1}}
\newcommand{\kfont}[1]{\textsc{#1}}

\newcommand{\power}[1]{\mathbb{P}({#1})}

\newcommand{\judgement}[5]{\ensuremath{{#1}: [#2]^{\textrm{\footnotesize $#3$}}_
{\textrm{\footnotesize {$#4$},{$#5$}}}}}
\newcommand{\tjudgement}[3]{\ensuremath{{#1} \downarrow_{\textrm{\footnotesize  $#3$}} {#2}}}

\newcommand{\mpre}[3]{\kfont{pre}({#1},{#2})={#3}}
\newcommand{\mpost}[3]{\kfont{post}({#1},{#2})={#3}}

\newtheorem{theorem}{Theorem}

\newenvironment{mrule}{\(\begin{array}{c}}
{\end{array}\)}

\newcommand{\ldbrack}{\Lbrack\,}
\newcommand{\rdbrack}{\,\Rbrack}
\newcommand{\ldangle}{\langle\hspace{-0.26em}\langle}
\newcommand{\rdangle}{\rangle\hspace{-0.26em}\rangle}

\newcommand{\semantics}[1]{\ensuremath{\ldbrack{#1}\rdbrack}}
\newcommand{\semanticsa}[1]{\ensuremath{\ldangle{#1}\rdangle}}

\newcommand{\tsemantics}[1]{\semanticsa{#1}}

\newcommand{\prevar}[1]{\ensuremath{\ffont{old}\ {#1}}}
\newcommand{\postvar}[1]{\ensuremath{\ffont{var}\ {#1}}}

\begin{document}
\maketitle

\begin{abstract}
We present an approach to program reasoning which inserts between a
program and its verification conditions an additional layer, the
denotation of the program expressed in a declarative form. The program
is first translated into its denotation from which subsequently the
verification conditions are generated. However, even before (and
independently of) any verification attempt, one may investigate the
denotation itself to get insight into the \enquote{semantic essence} of the
program, in particular to see whether the denotation indeed gives reason
to believe that the program has the expected behavior. Errors in the
program and in the meta-information may thus be detected and fixed prior
to actually performing the formal verification. More concretely,
following the relational approach to program semantics, we model the
effect of a program as a binary relation on program states. A formal
calculus is devised to derive from a program a logic formula that
describes this relation and is subject for inspection and manipulation.
We have implemented this idea in a comprehensive form in the RISC
ProgramExplorer, a new program reasoning environment for educational
purposes which encompasses the previously developed RISC ProofNavigator
as an interactive proving assistant.
\end{abstract}

\section{Introduction}

Most systems for program reasoning are based on calculi such as the Hoare
Calculus or Dynamic Logic~\cite{KeYBook2007} where we generate from a program
specification and a program implementation (which is annotated with additional
meta-information such as loop invariants and termination terms) those conditions
whose verification implies that the implementation indeed meets the
specification. The problem is that by such an approach we gain little insight
into the program before respectively independently of the verification process.
In particular, if the verification attempt is a priori doomed to fail because of
errors, inconsistencies, or weaknesses in the program's specification,
implementation, or meta-information (which is initially the case in virtually
all verification attempts), we will learn so only by unsuccessfully struggling
with the verification until some mental \enquote{click} occurs. This click
occurs frequently very late, because, in the heat of the struggle, it is usually
hard to see whether the inability to perform a correctness proof is due to an
inadequate proving strategy or due to errors or inconsistencies in the program.
Actually, it is usually the second factor that contributes most to the time
spent and frustration experienced; once we get the
specification/implementation/meta-information correct, the verification is a
comparatively small problem. We have frequently observed this fact in our own
verification attempts as well as in those performed by students of computer
science and mathematics in courses on formal methods.

We therefore advocate an alternative approach where we insert between a program
and its verification conditions an additional layer, the denotation of the
program~\cite{Schmidt1986} expressed in a declarative form. The program
(annotated with its meta-information) is translated into its denotation from
which subsequently the verification conditions are generated. However, even
before (and independently of) any verification attempt, one may investigate the
denotation itself to get insight into the \enquote{semantic essence} of the
program (independently of its \enquote{syntactic surface}), in particular to see
whether the denotation indeed gives reason to believe that the program has the
expected behavior. Errors in the program and in the meta-information may thus be
detected and fixed prior to actually performing the formal verification.

More concretely, following the relational approach to program
semantics~\cite{Lamport2002}, we model the effect of a program (command) $c$ as
a binary relation $\semantics{c}$ on program states which describes the possible
pairs of pre- and post-states of $c$. Such a relation can be also described in a
declarative form by a logic formula~$f_r$ with denotation $\semantics{f_r}$.
Thus a formal calculus is devised to derive from a program $c$ a judgment
$c:f_r$ such that $\semantics{c} \subseteq \semantics{f_r}$. For instance, we
can derive
\(
\pfont{x=x+1}: \postvar{x}=\prevar{x}+1
\)
where the logic variable $\prevar{x}$ refers to the value of the program
variable \pfont{x} in the prestate of the command and the logic variable
$\postvar{x}$ refers to its value in the poststate. In this way, we can
constrain the allowed state transitions, i.e.\ handle the partial correctness of
programs. To capture also total correctness, we introduce the set of states
$\tsemantics{c}$ on which the execution of $c$ must terminate ($\tsemantics{c}$
is a subset of the domain of $\semantics{c}$). Such a set can be also described
in a declarative form by a logic formula (a state condition)~$f_c$. Thus we
derive a judgment $c \downarrow f_c$ such that $\semantics{f_c} \subseteq
\tsemantics{c}$. In this fashion, the pair of formulas $f_r$ and $f_c$ captures
the semantic essence of $c$ in a declarative form that is open for inspection
and manipulation.

We have implemented this idea in a comprehensive form in the \emph{RISC
ProgramExplorer}\footnote{\url{http://www.risc.jku.at/research/formal/software/ProgramExplorer}},
a new program reasoning environment for educational purposes which encompasses
the previously developed \emph{RISC ProofNavigator} as an interactive proving
assistant~\cite{Schreiner2008}. The RISC ProgramExplorer supports reasoning
about programs written in a restricted form of Java (including support for
control flow interruptions such as \texttt{continue}, \texttt{break},
\texttt{return}, and \texttt{throw}, static and dynamic methods, classes and a
restricted form of objects) and specified in the formula language of the RISC
ProofNavigator (whose syntax is derived from PVS~\cite{PVS}). A
first version of the system has been released under the GNU
Public License in October 2011; it will be subsequently used in a regular course
on formal methods.
 
The remainder of this paper is structured as follows: Section~\ref{related}
discusses related work. Section~\ref{theory} sketches the theoretical
foundations of our approach, i.e.\ how programs are translated into their semantic
essence. Section~\ref{software} presents the implementation of this approach in
the RISC ProgramExplorer. Section~\ref{example} illustrates the use of the
software by a detailed example. Section~\ref{verifying} describes how from the
semantic essence the verification conditions are derived that show the
correctness of a program with respect to its specification. In
Section~\ref{workflow}, we describe the typical workflow supported by the
system; in Section~\ref{conclusions}, we conclude and discuss further work.

\section{Related Work}
\label{related}

There exist numerous frameworks and tools for reasoning about programs written
in various languages. In the context of the programming language Java, the Java
Modeling Language (JML) has become the de facto standard specification
language~\cite{JML}; it extends the syntax and semantics of the Java expression
language to a mathematical formula language that is rich enough to formulate
method contracts. Various tools aim at reasoning about JML-annotated programs in
a fully automatic way, mainly in order to falsify programs (find runtime errors
or violations of method preconditions) rather than to verify them. An
environment for true verification is the system Why/Krakatoa~\cite{FilliatreM07}
which translates JML-annotated programs into the input language of the
verification condition generator Why that produces verification conditions for a
variety of external provers such as the interactive proving assistant Coq.
Likewise, the formal software development tool KeY~\cite{KeYBook2007} allows to
verify JML-specified programs written in JavaCard (a subset of Java) based on
the framework of dynamic logic (a modal logic whose modalities integrate program
statements) using a built-in interactive prover. In contrast to these
approaches, the approach presented in this paper is based on a semantic model
that is visible for human inspection, reasoning is essentially based on
classical predicate logic and on a formula/specification language that is
independent of any programming language; the supported programming language
has also Java-like syntax but is not fully object-oriented 
(currently no inheritance is supported).

The idea to translate computer programs to mathematical objects that describe
the meaning of the programs originates from the work of Scott and Strachey in
the 1960s and is now known under the name \enquote{denotational semantics}. In
\cite{Schmidt1986}, this approach is presented as a methodology for language
design which also helps to understand concrete programs by investigating the
mathematical objects to which they are translated. In the classical
Scott/Strachey approach, programs are translated to functions; loops are
translated to recursive function equations over semantic domains with special
properties (pointed complete partial orders) that ensure the existence and
uniqueness of the function (as the least fixed point of the defining equation).
In contrast to the classical approach, in our model programs are translated to
relations described by formulas that do not necessarily imply existence and
uniqueness of the result (as the formula is in general derived from
user-provided method contracts and loop invariants).

The idea of modeling programs as state relations is not new; it has emerged in
various approaches to the formalization of programs: it is for example the core
idea of the Lamport's \enquote{Temporal Logical of Actions} (TLA)
\cite{Lamport2002} where the individual actions of a process are described by
formulas relating pre- to post-states; the language PlusCal for the formulation
of algorithms is translated to TLA specifications. Boute's
\enquote{Calculational Semantics} \cite{Boute} defines the behavior of a program
by equations relating the pre-state to the post-state. Hoare and
Jifeng's \enquote{Unifying Theories of Programming}~\cite{HoareJifeng} provides an integrated theory of programming based on a view of programs as relations.
Apart from TLA (which provides a model checker for a final state
subset of the specification language), we are not aware of any software tools or
verification environments that are directly based on relational frameworks.

Also the principle of \enquote{correct by construction}, which emphasizes the
gradual refinement of a specification into an executable program such that at
any stage of the refinement the program is provably correct, is usually based on
a relational view of programs. This approach was originally pioneered by
Dijkstra~\cite{Dijkstra1968} and further elaborated by Back's
\enquote{refinement calculus}~\cite{Back} and other formalisms of a similar
flavor~\cite{Back,Morris,Morgan,Hehner,Gordon}. In some way or another, these
calculi allow declarative specifications to reside on the same language level as
operational commands; the endorsed approach to program development is to
gradually transform specifications to commands. %Also the JML feature of
%\enquote{model programs} is based on the refinement calculus. 

A recent variant of this principle is Back's \enquote{invariant-based
programming}~\cite{Back2009} where program invariants are constructed before the
actual code is considered. This approach is implemented in the Socos environment
which provides a graphical editor for the construction of invariant diagrams
from which verification conditions are generated for the PVS prover. Although
not directly based on these principles, also the approach presented in this
paper can be applied in a top-down refinement fashion: since the transition
relation of a program that executes a loop or calls a method is only constructed
from the invariant of the loop respectively specification of the method, we may
verify the correctness of the program before the body of the loop respectively
method is implemented. 

The appropriate way (didactic approach, formal calculus, tool support) of
teaching formal methods is an ongoing point of debate, see e.g.\ the TFM
conference series~\cite{TFM2004,TFM2006,TFM2009}. A comparative survey of formal
methods courses in Europe is given in~\cite{Oliveira2004}; a comparison of tools
for teaching program verification is presented in~\cite{Feinerer2009}. In
general, most knowledge on formal methods education is based on
personal experience reports; there is hardly any scientific evidence for the
superiority of any particular approach.

%Likewise, by the use of assertions, we may define for certain
%program points desired post-state knowledge that needs to be established for the
%correct execution of the program; later actual code may be provided that
%transforms the program point's pre-state into the desired post-state.

\section{Commands as State Relations}
\label{theory}

We base the presentation of our formalism on a simple command language without control flow interruptions and method calls. In this language, a command $c$ can be formed according to the grammar
\begin{quote}
$c ::= x\ \pfont{=}\ e\ |\
\pfont{\{var}\ x\pfont{;}\ c\pfont{\}}\ |\
\pfont{\{}c_1\pfont{;}c_2\pfont{\}}\ |\
\pfont{if}\ \pfont{(}e\pfont{)}\ \pfont{then}\ c\ |\
\pfont{if}\ \pfont{(}e\pfont{)}\ \pfont{then}\ c_1\ \pfont{else}\ c_2\ |\
\pfont{while}\ \pfont{(}e\pfont{)}^{f,t}\ c$
\end{quote}
where $x$ denotes a program variable, $e$ denotes a program expression, and a while loop is annotated by an invariant formula $f$ and termination term $t$.
The semantics of a command $c$ is defined, for a given set $\mathit{Store}$ of possible states (store contents), by a binary relation $\semantics{c}\subseteq\mathit{Store}\times\mathit{Store}$ that defines the possible state transitions of the command and by a set $\tsemantics{c}\subseteq\mathit{Store}$ that defines those pre-states where the command must perform a transition to some post-state; for a definition of the semantics, see~\cite{Schreiner2008c}.

In Figures~\ref{transition}, \ref{termination}, and~\ref{prepost}, we give rules
(where the terms \prevar{xs} and \postvar{xs} refer to the sets of values of the program variables $\mathit{xs}$ in the pre-/post-state) to derive 
the following three kinds of judgments:

\begin{figure}[p]
\begin{center}
\hfill
\begin{mrule}
\hfill
\judgement{c}{f}{\mathit{xs}}{g}{h} \hfill
x \not\in \mathit{xs} \hfill \\
\hline
\judgement{c}{f \wedge \postvar{x}=\prevar{x} }{\mathit{xs}\;\cup\;\{\mathit{x}\}}{g}{h}
\end{mrule}
\hfill
\begin{mrule}
e \simeq_{h} t \\
\hline
\judgement{x\ \pfont{=}\ e}{\postvar{x}=t}{\{x\}}{\ffont{true}}{h}
\end{mrule}
\hfill
\medskip

\begin{mrule}
\judgement{c}{f}{\mathit{xs}}{g}{h} \\
\hline
\judgement{\pfont{\{var}\ x\pfont{;}\ c\pfont{\}}}
{\exists x_0,x_1: f[x_0/\prevar{x},x_1/\postvar{x}]}
{\mathit{xs}\backslash x}{g}{\forall x: \mathit{h}[x/\prevar{x}]}
\end{mrule}
\medskip

\begin{mrule}
\hfill
\judgement{c_1}{f_1}{\mathit{xs}}{g_1}{h_1} \hfill
\judgement{c_2}{f_2}{\mathit{xs}}{g_2}{h_2} \hfill 
\mpre{c_1}{h_2}{h_3} \hfill \\
\hline
\judgement{\pfont{\{}c_1\pfont{;}c_2\pfont{\}}}{\exists\mathit{ys}:
f_1[\mathit{ys}/\postvar{\mathit{xs}}] \wedge
f_2[\mathit{ys}/\prevar{\mathit{xs}}]}{\mathit{xs}}{g_1\wedge g_2}{h_1 
\wedge h_3}
\end{mrule}
\medskip

\begin{mrule}
\hfill
e \simeq_h f_e \hfill
\judgement{c_1}{f_1}{\mathit{xs}}{g_1}{h_1} \hfill \\
\hline
\judgement{\pfont{if}\ \pfont{(}e\pfont{)}\ \pfont{then}\ c}
{\ffont{if}\ f_e\ \ffont{then}\ f_1\ \ffont{else}\ \postvar{\mathit{xs}}=\prevar{\mathit{xs}}}{\mathit{xs}}{g_1}{h \wedge (f_e \Rightarrow h_1)}
\end{mrule}
\medskip

\begin{mrule}
\hfill
e \simeq_h f_e \hfill
\judgement{c_1}{f_1}{\mathit{xs}}{g_1}{h_1} \hfill 
\judgement{c_2}{f_2}{\mathit{xs}}{g_2}{h_2} \hfill \\
\hline
\judgement{\pfont{if}\ \pfont{(}e\pfont{)}\ \pfont{then}\ c_1\ \pfont{else}\ c_2}
{\ffont{if}\ f_e\ \ffont{then}\ f_1\ \ffont{else}\ f_2}{\mathit{xs}}{g_1 \wedge g_2}{h \wedge \ffont{if}\ f_e\ \ffont{then}\ h_1\ \ffont{else}\ h_2}
\end{mrule}
\medskip

\begin{mrule}
\hfill
e \simeq_{h} f_e \hfill 
\judgement{c}{f_c}{\mathit{xs}}{g_c}{h_c} \hfill \\
g \equiv
\forall \mathit{xs},\mathit{ys},\mathit{zs}:
\mathit{f}[\mathit{xs}/\prevar{xs},\mathit{ys}/\postvar{xs}] \wedge f_e[\mathit{ys}/\prevar{xs}]
\wedge f_c[\mathit{ys}/\prevar{xs},\mathit{zs}/\postvar{xs}] 
\Rightarrow \\
h[\mathit{ys}/\prevar{xs}] \wedge
f[\mathit{xs}/\prevar{xs},\mathit{zs}/\postvar{xs}] \\
\hline
\judgement{\pfont{while}\ \pfont{(}e\pfont{)}^{f,t}\ c}{f \wedge \neg f_e[\postvar{xs}/\prevar{xs}]}
{\mathit{xs}}{g_c \wedge g}{h \wedge f[{\prevar{xs}/\postvar{xs}}]}
\end{mrule}

\end{center}
\caption{The Transition Rules}
\label{transition}
\end{figure}

\begin{figure}
\begin{center}
\hfill
\begin{mrule}
\tjudgement{x\ \pfont{=}\ e}{\ffont{true}}{\ffont{true}}
\end{mrule}
\hfill
\begin{mrule}
\tjudgement{c}{f}{g} \\
\hline
\tjudgement{\pfont{\{var}\ x\pfont{;}\ c\pfont{\}}}
{\forall x: f[x/\prevar{x}]}{g}
\end{mrule}
\hfill
\begin{mrule}
\hfill
\tjudgement{c_1}{f_1}{g_1}\hspace{1em} \hfill
\tjudgement{c_2}{f_2}{g_2}\hspace{1em} \hfill 
\mpre{c_1}{f_2}{f_3} \hfill \\
\hline
\tjudgement{\pfont{\{}c_1\pfont{;}c_2\pfont{\}}}{f_1 \wedge f_3}{g_1 \wedge g_2}
\end{mrule}
\hfill
\medskip

\hfill
\begin{mrule}
\hfill
e \simeq_h f_e \hfill
\tjudgement{c}{f}{g} \hfill \\
\hline
\tjudgement{\pfont{if}\ \pfont{(}e\pfont{)}\ \pfont{then}\ c}
{f_e\ \Rightarrow f}{g}
\end{mrule}
\hfill
\begin{mrule}
\hfill
e \simeq_h f_e \hfill
\tjudgement{c_1}{f_1}{g_1} \hfill 
\tjudgement{c_2}{f_2}{g_2} \hfill \\
\hline
\tjudgement{\pfont{if}\ \pfont{(}e\pfont{)}\ \pfont{then}\ c_1\ \pfont{else}\ c_2}
{\ffont{if}\ f_e\ \ffont{then}\ f_1\ \ffont{else}\ f_2}
{g_1 \wedge g_2}
\end{mrule}
\hfill
\medskip

\begin{mrule}
\hfill
e \simeq_{h} f_e \hfill 
\judgement{c}{f_c}{\mathit{xs}}{g_c}{h_c} \hfill 
\tjudgement{c}{f_t}{g_t} \hfill \\
g \equiv
\forall \mathit{xs},\mathit{ys},\mathit{zs}:
\mathit{f}[\mathit{xs}/\prevar{xs},\mathit{ys}/\postvar{xs}] \wedge f_e[\mathit{ys}/\prevar{xs}]
\wedge f_c[\mathit{ys}/\prevar{xs},\mathit{zs}/\postvar{xs}] 
\Rightarrow \\
f_t[\mathit{ys}/\prevar{xs}] \wedge
0 \leq t[\mathit{zs}/\prevar{xs}] < t[\mathit{ys}/\prevar{xs}]\ \\
\hline
\tjudgement{\pfont{while}\ \pfont{(}e\pfont{)}^{f,t}\ c}{t >= 0}{g \wedge g_t}
\end{mrule}
\end{center}
\caption{The Termination Rules}
\label{termination}
\end{figure}

\begin{figure}
\begin{center}
\begin{mrule}
\judgement{c}{f}{\mathit{xs}}{g}{h} \\
\hline
\mpre{c}{f_q}
{\forall \mathit{xs}: f[\mathit{xs}/\postvar{xs}] \Rightarrow f_q[\mathit{xs}/\prevar{xs}]}
\end{mrule}
\medskip

\begin{mrule}
\judgement{c}{f}{\mathit{xs}}{g}{h} \\
\hline
\mpost{c}{f_p}
{\exists \mathit{xs}: f_p[\mathit{xs}/\prevar{xs}] \wedge f[\mathit{xs}/\prevar{xs},\prevar{xs}/\postvar{xs}]}
\end{mrule}
\end{center}
\caption{The Pre-/Postcondition Rules}
\label{prepost}
\end{figure}

\begin{itemize}
\item $\judgement{c}{f_r}{\mathit{xs}}{g}{h}$ denotes the derivation of a state relation $f_r$ from command $c$ together with the set of program variables $\mathit{xs}$ that may be modified by $c$. The remaining arguments $g$ and $h$ express additional side-conditions (which
may be ignored on first reading): the derived relation is correct if
the derived state-independent condition $g$ holds, and if 
the derived state condition $h$ holds on the pre-state of~$c$.
The rationale for $g$ is to capture state-independent conditions such as the
correctness of loop invariants; the purpose of $h$ is to capture statement
preconditions that prevent e.g.\ arithmetic overflows. These side conditions
have to be proved; they are separated from the transition relation~$f_r$ to make
the core of the relation better understandable. 
\item $\tjudgement{c}{f_c}{g_c}$ denotes the derivation of a state condition
(termination condition) $f_c$ from $c$; the derived condition is correct, if the
state-independent condition $g_c$ holds. The purpose of this side condition is
to capture that the execution of every loop body terminates and decreases
the value of the termination term but does not make the value negative.
\item $\mpre{c}{f_q}{f_p}$ and $\mpost{c}{f_p}{f_q}$ denote derivations that
compute from a command $c$ and a condition $f_q$ on the post-state of $c$ a
corresponding condition $f_p$ on the pre-state, respectively from~$c$ and
pre-condition~$f_p$ the post-condition $f_q$. The corresponding rules in
Figure~\ref{prepost} show that these conditions can be computed directly from
the transition relation of $c$.
\end{itemize}

The derivations use additional judgments $e \simeq_{f_e} f$ and $e
\simeq_{f_e} t$ which translate a Boolean-valued program expression $e$ into a
logic formula $f$ and an expression $e$ of any other type into a term $t$,
provided that the state in which $e$ is evaluated satisfies the condition $f_e$
(the rules for these judgments are omitted).

One should note that the rules presented in Figures~\ref{transition} and~\ref{termination} can be applied recursively over the structure of a
command; first we determine the transition/termination formula of the
subcommands, then we combine the formulas to the transition/termination formula
of the whole command. Along this process, the side condition $h$ is constructed which has to be shown separately to hold in the pre-state of the command in order to verify the correctness of the translation.

A special case is the rule for \pfont{while} loops. Here the result is only
determined from the invariant formula respectively termination term by which the
loop is annotated; additionally, a proof obligation~$g$ is generated to verify
the correctness of the loop body with respect to invariant and termination term.
In a similar way, in the full programming language calls of program methods are
handled: the transition relation of the method call is derived from the
specification of the method; the correctness of the implementation of the method
is to be established separately. We thus yield a modular approach to the
derivation of transition relations and termination conditions; the size of the
derived formula is independent of the sizes of the bodies of the loops executed
respectively of the methods called. Furthermore, as already stated in
Section~\ref{related}, the approach gives rise to some sort of \enquote{correct
by construction} approach: we may first develop loop invariants and method
preconditions (and verify the correctness of programs executing the loops and
calling the methods) before we implement the bodies of the loops and methods
(and consequently verify the correctness of implementations).

Formally, the derivations satisfy the following soundness constraints.
\begin{theorem}[Soundness]
For all $c\in\mathit{Command}, f_r,f_c,f_p,f_q,g,h\in\mathit{Formula}, \mathit{xs}\in\power{Variable}$, the following statements hold:
\begin{enumerate}
\item If we can derive the judgment
$\judgement{c}{f_r}{\mathit{xs}}{g}{h}$,
then we have for all $s,s'\in\mathit{Store}$
\begin{displaymath}
\semantics{g} \wedge \semantics{\mathit{h}}(s) \Rightarrow
(\semantics{c}(s,s') \Rightarrow
\semantics{f_r}(s,s') \wedge \forall x\in\mathit{Variable}\backslash\mathit{xs}: \semantics{\mathit{x}}(s)=\semantics{x}(s')).
\end{displaymath}
\item If we can (in addition to $\judgement{c}{f_r}{\mathit{xs}}{g}{h}$) 
derive the judgment
$\tjudgement{c}{f_c}{g_c}$,
then we have for all $s\in\mathit{Store}$
\begin{displaymath}
\semantics{g} \wedge \semantics{g_c} \wedge
\semantics{\mathit{h}}(s) \Rightarrow
(\semantics{f_c}(s) \Rightarrow
\tsemantics{c}(s)).
\end{displaymath}
\item If we can (in addition to $\judgement{c}{f_r}{\mathit{xs}}{g}{h}$) also derive the judgment $\mpre{c}{f_q}{f_p}$ or 
the judgment $\mpost{c}{f_p}{f_q}$, then we have for all $s,s'\in\mathit{Store}$
\begin{displaymath}
\semantics{g} \wedge \semantics{\mathit{h}}(s) \Rightarrow
(\semantics{f_p}(s) \wedge
\semantics{f_r}(s,s') \Rightarrow \semantics{f_q}(s')).
\end{displaymath}
\end{enumerate}
\end{theorem}
The semantics $\semantics{f}(s,s')$ of a transition relation $f$ is determined over a pair of states $s,s'$ (and a logic environment, which is omitted for clarity); the semantics of state condition $g$ is defined as $\semantics{g}(s)\Leftrightarrow \forall s': \semantics{g}(s,s')$ and the semantics of a state independent-condition $h$ is defined as $\semantics{h}\Leftrightarrow \forall s,s': \semantics{h}(s,s')$.

In~\cite{Schreiner2008b}, the formal semantics of commands and formulas has been
defined and the soundness of (a preliminary form of) the calculus has been
proved. In~\cite{Schreiner2008c}, a concise definition of the semantics,
judgments, and rules of (a preliminary form of) the calculus is given.

\section{The RISC ProgramExplorer}
\label{software}

\begin{figure}
\begin{center}
\includegraphics[width=0.8\textwidth]{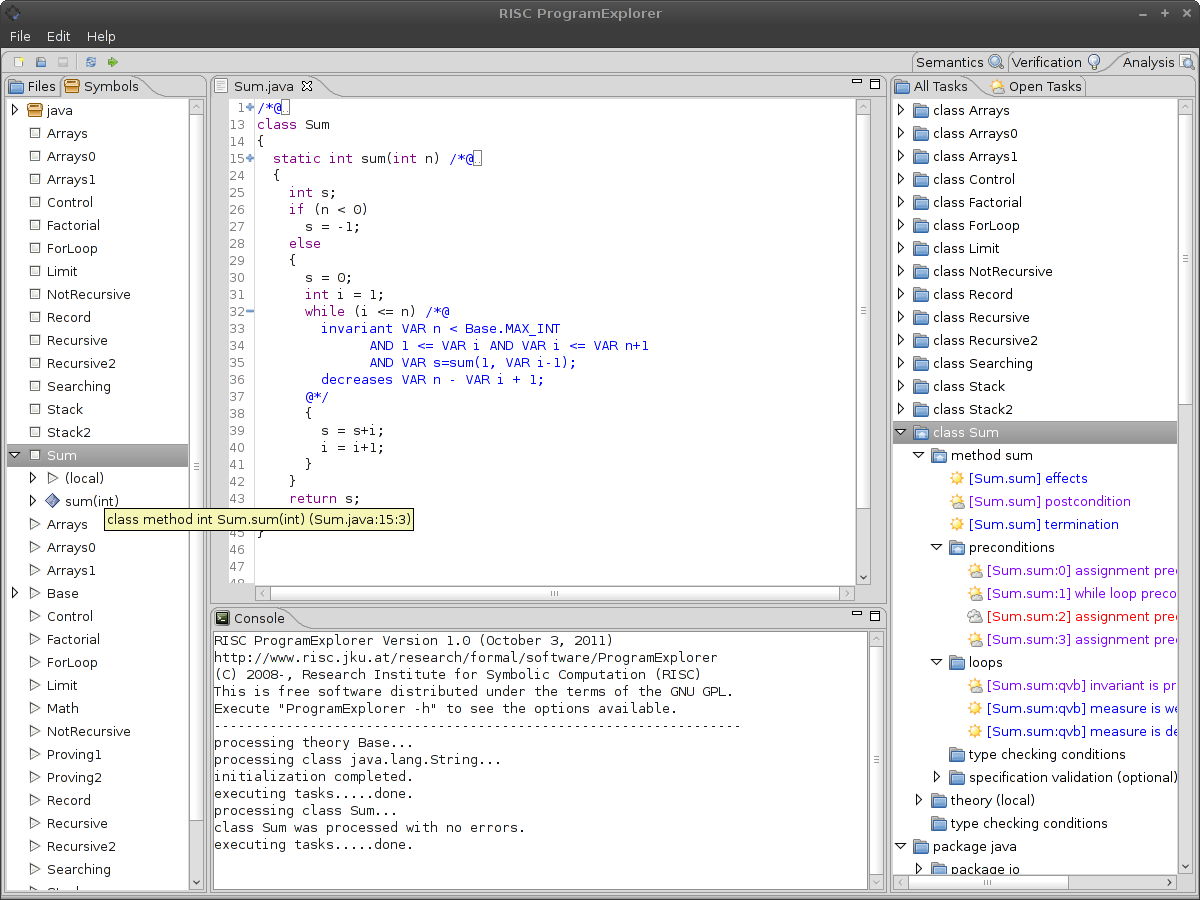}
\end{center}
\caption{The RISC ProgramExplorer (Analysis View)}
\label{explorer}
\end{figure}

We have implemented the calculus presented in the previous section in the RISC ProgramExplorer; a screenshot of the software is given in Figure~\ref{explorer}. 

The RISC ProgramExplorer supports reasoning about programs written in a subset
of Java which we call \enquote{MiniJava}. This subset includes classes with
class and object variables as well as class and object methods and constructors.
Method bodies may execute most kinds of Java commands including those that cause
interruptions of the control flow (\pfont{continue}, \pfont{break},
\pfont{return}, \pfont{throw}). The major restriction compared with full Java is
that the type checker prevents sharing of objects/arrays by different variables
(such that data structures can be modeled as plain values rather than as pointer
structures) and that inheritance is not supported; furthermore expressions are
not allowed to cause side effects (thus method calls with result values have to
be written as separate commands). 

The software supports a theory definition language that is derived from the
language of the previously developed RISC ProofNavigator~\cite{Schreiner2008};
the syntax of the language is inherited from PVS~\cite{PVS} respectively CVC
Lite~\cite{CVCL}. The language allows to introduce new theories consisting of
types, objects, functions, and predicates as well as axioms and theorems (which
can be proved with the RISC ProofNavigator). Theories may be specified in
separate files for reuse in different programs or may be attached to the program
classes where they are used.

Based on these theories, programs may be formally specified by class invariants,
method contracts, loop invariants, termination terms, and assertions in the
style of the Java Modeling Language~\cite{JML}. The syntax and semantics of the
formula language is, however, deliberately neutral of the programming language
(because we believe that mathematics precedes programming and the same
mathematical definitions and specifications should be reusable for different
programming languages). The specifications operate therefore directly on the
semantic model of a program (e.g.\ every program class $C$ is translated to a
theory $C$ that contains among other definitions a record type $C$; if $x$ is a
program variable that denotes an object of class $C$, then in a specification
$\ffont{var}\ x$ denotes a record of type $C$).

The RISC ProgramExplorer provides an elaborated graphical user interface 
that links theories, programs, semantic models, and
verification tasks by three main views:
\begin{description}
\item[Analysis] This is the central view in which mathematical theories and
program classes may be developed (see Figure~\ref{explorer}); upon saving the
corresponding file, the theory respectively program is type-checked, the
semantic model is constructed, and verification tasks are generated. If an error
occurs during this semantic processing, the error is linked to the corresponding
location in the source code; likewise the generated tasks are linked to
the corresponding source locations. In detail, the following verification tasks
are generated:
\begin{description}
\item[Effects] The proof that the method does not modify any variable outside the specified set of variables and does not throw any unspecified exception.
\item[Postcondition] The proof that the transition relation derived from
the body of a method implies the method's postcondition.
\item[Termination] The proof that a method's termination condition 
implies the termination condition derived from the body of the method.
\item[Preconditions] The proofs that every statement is only executed in a state
that satisfies the precondition of the statement.
\item[Loops] The proofs that a loop body preserves the invariant, that the
execution of the body terminates and 
decreases the value of the specified termination term, and that the
decreased value does not become negative. 
\item[Type checking conditions] The proofs that all formulas are well-typed (not all type checking questions of the formula language can be statically answered).
\item[Specification validation] The (optional) proofs that a specification is satisfiable (for every argument that satisfies the precondition there is a result that satisfies the postcondition) but not trivially satisfiable (there is also a result that does not satisfy the postcondition).
\end{description}
Splitting the overall task of proving the correctness of a method into individual
subtasks supports the gradual verification of different aspects of correctness
and gives more concrete hints in the case of failed proof attempts. Some of the
tasks can be fully automatically solved by the validity checker; if the checker
fails, the user can start a semi-automatic interactive proof.
\begin{figure}[t]
\begin{center}
\includegraphics[width=0.75\textwidth]{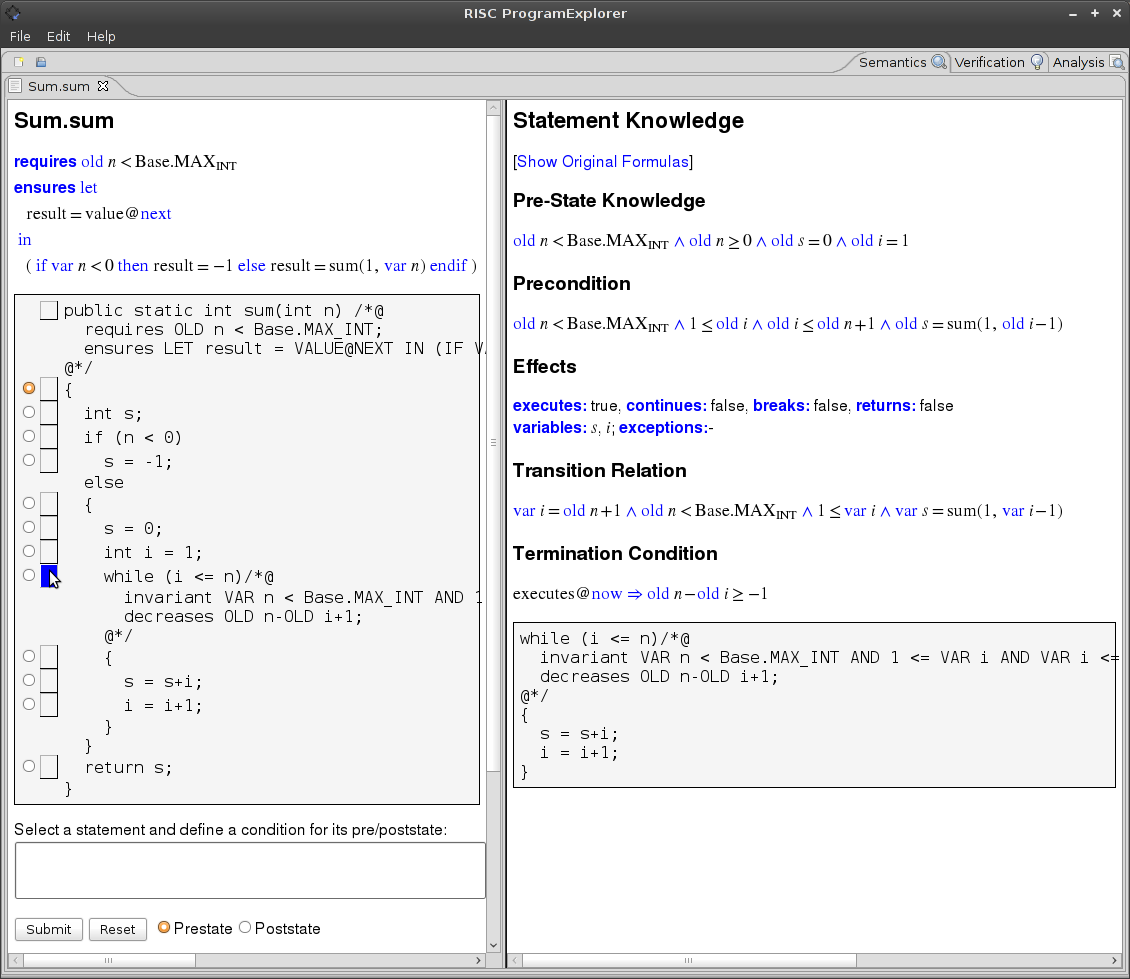}
\end{center}
\caption{Semantics View}
\label{semantics}
\end{figure}
\item[Semantics] In this view displayed in Figure~\ref{semantics}, the semantic model of a selected program method may be investigated. By moving the mouse pointer, the user may display for each command of the method body (respectively for the whole body)

\begin{figure}
\begin{center}
\includegraphics[width=0.75\textwidth]{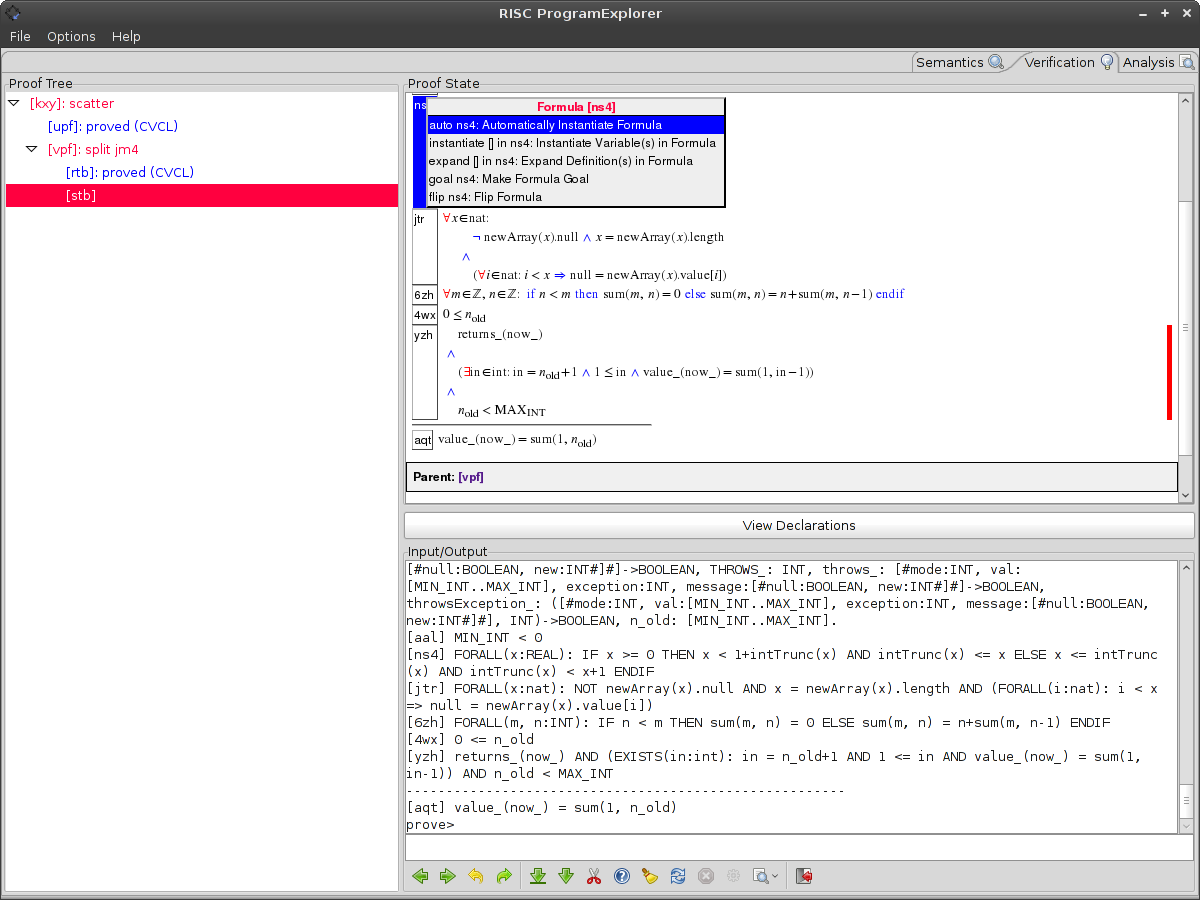}
\end{center}
\caption{Verification View}
\label{verification}
\end{figure}

\begin{itemize}
\item the transition relation of the command,
\item the termination condition of the command,
\item the effect of the command (i.e.\ the set of variables potentially changed,
the set of exceptions potentially thrown, the information whether the command
may interrupt the control flow by executing a \pfont{continue}, \pfont{break},
\pfont{return} statement),
\item the precondition of the command,
\item the condition which is known to hold on the pre-state of the command.
\end{itemize}
Furthermore, the user may enter a desired condition on the pre-/post-state of a
selected command; the system then determines the
consequences for the pre-states of all other commands.
\item[Verification] In this view, the user may perform a semi-interactive proof
of a selected verification task with the help of the RISC ProofNavigator which
is embedded in the RISC ProgramExplorer. This proving assistant provides a small
set of commands that implement typical proving strategies on the level of
formulas (term reasoning is delegated to the validity checker CVCL~\cite{CVCL});
the most frequently used commands are bound to buttons respectively menus
attached to corresponding formulas. By the use of these commands, proof
situations are gradually reduced to sub-situations with all proof situations
displayed in a tree structure; the goal is to achieve such situations that can
be automatically determined as valid by the validity checker. The interface has
been carefully designed to make the handling of proofs as convenient as
possible.

Proofs respectively proof attempts are persistently stored on disk and can be
later replayed; the status of a proof and the dependencies of proofs to theory
definitions or separately proved formulas are automatically managed (it is e.g.\
indicated whether a previously performed proof is still valid, i.e.\ whether no
prerequisite of the proof has changed).

Figure~\ref{verification} displays the verification view; for details of the
RISC ProofNavigator, see~\cite{Schreiner2008}.

\end{description}

\section{An Example}
\label{example}

We illustrate the use of the RISC ProgramExplorer by the following method
\pfont{sum} which returns, for non-negative argument $n$, the sum of all integers
from $1$ to $n$, and for negative $n$, the value $-1$: 

\begin{footnotesize}
\begin{verbatim}
  static int sum(int n) /*@
    requires VAR n < Base.MAX_INT;
    ensures 
      LET result=VALUE@NEXT IN
      IF VAR n < 0 
        THEN result = -1 
        ELSE result = sum(1, VAR n) 
      ENDIF;
  @*/
  {
    int s;
    if (n < 0)
      s = -1;
    else {
      s = 0; 
      int i = 1;
      while (i <= n) /*@
        invariant VAR n < Base.MAX_INT 
              AND 1 <= VAR i AND VAR i <= VAR n+1 
              AND VAR s=sum(1, VAR i-1);
        decreases VAR n - VAR i + 1;
      @*/ 
      {
        s = s+i; 
        i = i+1;
      }
    }
    return s;
  }
\end{verbatim}
\end{footnotesize}
The method is specified by a pair of pre- and postcondition; the term
\ffont{value@next} in the postcondition refers to the return value of the
function; the program type \pfont{int} is mapped to the specification type
\ffont{Base.int} which denotes the set of all integers from
\ffont{Base.MIN\_INT} to \ffont{Base.MAX\_INT}. One should note that the
method is actually not completely correct with respect to the specification,
since the computation of result $s$ might yield an overflow 
(see the end of this section for more information on this aspect).

The specification uses a binary function $\mathsf{sum}: \mathbb{Z}
\times \mathbb{Z} \rightarrow \mathbb{Z}$ such that $\mathsf{sum}(m,n)$ denotes
the sum of all integers from $m$ to $n$; this function is specified in a
theory which can be attached to the class in which the method is located:

\begin{footnotesize}
\begin{verbatim}
  theory {
    sum: (INT, INT) -> INT;
    sumaxiom: AXIOM
      FORALL(m: INT, n: INT):
        IF n<m 
          THEN sum(m, n) = 0
          ELSE sum(m, n) = n+sum(m, n-1)
        ENDIF;
  }
\end{verbatim}
\end{footnotesize}
Here \textsf{INT} denotes the set of all integers; 
$\mathsf{sum}$ is axiomatized rather
than defined, because the theory language does currently
not support recursive definitions. 

Based on the calculus presented in Section~\ref{theory}, the RISC Program
Explorer translates the while loop to the following semantic model (the loop's
precondition and pre-state knowledge are not shown):
\begin{quote}
\includegraphics[width=10cm]{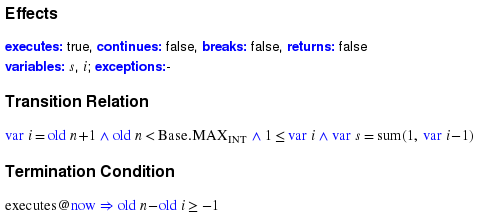}
\end{quote}
Here the core of the transition relation is the formula
$\postvar{i}=\prevar{n}+1\wedge\postvar{s}=\mathsf{sum}(1,\postvar{i}-1)$ (which
implies $\postvar{s}=\mathsf{sum}(1,\prevar{n})$) while the core of termination
condition is $\prevar{n}-\prevar{i}\geq -1$ (the initial value of the
termination term must not be negative). The formulas derived by the plain
calculus are actually (in general) much more complex; the human-friendly form
shown above is derived only after performing extensive processing by a built-in
simplifier (the user may freely switch between the display of the simplified
formula and its original version).

From this translation, the conditional statement is translated as follows (the
termination condition is automatically simplified to \ffont{true} and therefore
not displayed any more):
\begin{quote}
\includegraphics[width=8cm]{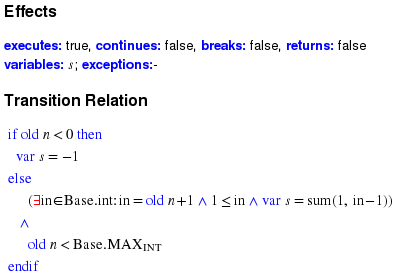}
\end{quote}
The whole body of the method is translated to
\begin{quote}
\includegraphics[width=8.5cm]{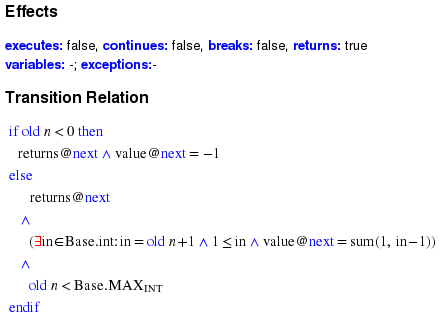}
\end{quote}

Here the \enquote{effects} clause indicates that the body of the method results
in the execution of a \pfont{return} statement, which is also indicated by the
formula \ffont{returns@next} in the transition relation. The core of the
transition relation in the second branch is $\exists
\mathit{in}:\mathit{in}=\prevar{n}+1 \wedge
\mathit{value}\ffont{@next}=\mathsf{sum}(1,\mathit{in}-1)$ which implies
$\mathit{value}\ffont{@next}=\mathsf{sum}(1,\prevar{n})$. The transition
relation denotes the semantic essence of the method which concisely describes
the behavior of the method in a declarative form; from this, the correctness of
the method according to its specification is quite self-evident even before the
formal proof is started.

As a side effect of the translation, the RISC ProgramExplorer
generates a couple of verification tasks, which will be explained in more detail
in Section~\ref{verifying}. One of them is the obligation to prove that the
postcondition $q$ is implied by the precondition $p$ and the derived transition
relation $r$ (i.e.\ to prove $p \wedge r \Rightarrow q$ as described in
Section~\ref{verifying}):
\smallskip

\begin{quote}
\includegraphics[width=10.5cm]{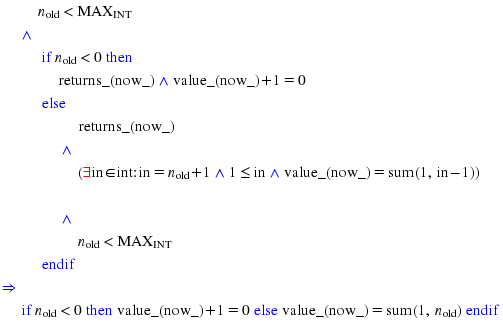}
\end{quote}
The proof of this formula proceeds by executing three commands, two triggered by
pressing a button, one by selecting a command from a formula menu
(see~\cite{Schreiner2008} for a more detailed description of the interaction
with the prover).
The correspondingly generated proof tree (whose nodes are labeled by
the respective proof commands) is

\begin{quote}
\includegraphics[width=4.5cm]{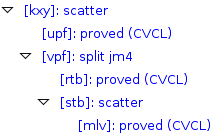}
\end{quote}
This proof would be irrelevant, if the post-condition $q$ specified by the user
were \enquote{trivial}, i.e., satisfied by every output value (generally
indicating an error in the specification). To rule this out, an (optional)
verification condition is generated that validates the specification by showing 
that, for every prestate that satisfies precondition $p$, there exists a 
poststate that does \emph{not} satisfy $q$ (i.e., $\forall x: p(x) 
\Rightarrow \exists y: \neg q(x,y)$ as described in Section~\ref{verifying}):
\smallskip

\begin{quote}
\includegraphics[width=10cm]{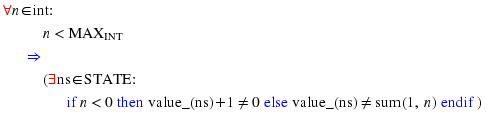}
\end{quote}
The corresponding proof proceeds (as determined by the branch condition
\pfont{n<0} in the program) by manual case distinction and, in each case, by an
instantiation with a state that holds an incorrect return value (there also
arises a third case, because we have declared the theory function
\ffont{sum} over the integers, such that in principle a negative result might
arise):
\smallskip

\begin{quote}
\includegraphics[width=8cm]{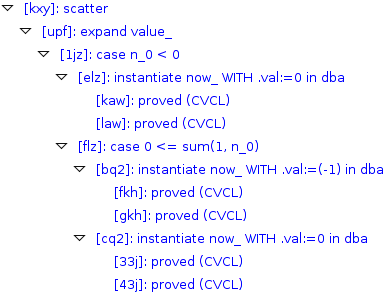}
\end{quote}

Another core verification task is the proof that the loop body with transition
relation $r$ preserves the invariant $i$ (i.e.\ the proof of $i' \wedge e \wedge
r \Rightarrow i''$ as described in Section~\ref{verifying}). After the built-in 
logical simplification, the proof goal becomes
\smallskip

\begin{quote}
\includegraphics[width=7.5cm]{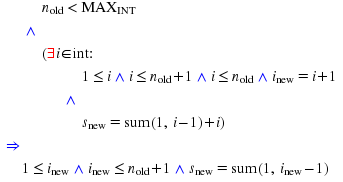}
\end{quote}
The corresponding proof is performed by two commands, one of which is the manual
instantiation of the axiom defining the function $\ffont{sum}$; the
corresponding proof tree is
\smallskip

\begin{quote}
\includegraphics[width=6cm]{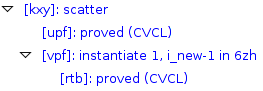}
\end{quote}

The proofs of the other verification conditions proceed mostly automatically;
the only exception is the proof that the increment \pfont{s = s+i} does not
yield an overflow. As already stated in the beginning of this section, this
is actually not true for our method. To make it true, the specification has
to be extended by an additional precondition that puts an upper limit on the
value of the sum (\pfont{sum(1,VAR n) <= Base.MAX\_INT}); to perform
the corresponding verification, then some additional lemmas about the 
monotonicity of $\ffont{sum}$ have to be introduced and proved.

The example presented in this section only involves simple control structures
and uses integer numbers as the only datatype. However, the full calculus also
supports programs with commands interrupting the control flow; these are
translated to formulas that involve special atomic predicates to express the
behavior of a program with respect to control flow (e.g.\ the predicate
\ffont{returns@\textit{state}} shown above indicates that \emph{state}
results from the execution of a \pfont{return} statement). Furthermore, the
implementation supports the usual datatypes like arrays and objects. The transition relations derived from programs involving these
features are apparently more complex than the examples shown above;
nevertheless they still become manageable after appropriate simplification. For
this purpose, it is recommended to express method specifications and loop
invariants with the help of high-level functions and predicates introduced in
theory declarations. The derived transition relations then also refer to these
formulas and become (again, after appropriate simplification) essentially as
readable as the specifications and invariants. The distribution of the RISC
ProgramExplorer comes with a couple of examples (e.g.\ operations on arrays like
sorting, linear search, binary search, a class implementing the datatype
\enquote{stack}, and others) that may serve as a starting point for further
applications.

\section{State Relations and Verification Conditions}
\label{verifying}

One of the advantages of the translation of a program into its \enquote{semantic
essence}, i.e., into a formula denoting its state relation, is that by this 
translation the derivation and interpretation of the individual verification
tasks becomes very transparent. In the following we explain the logical
interpretation of the various verification tasks generated by the 
RISC ProgramExplorer from every method and their relationship to 
the method's semantic essence.

\paragraph{Postcondition} For a method with precondition $p$, postcondition $q$ 
and state relation $r$ of the method body, the goal of this task is to prove
\begin{quote}
  $p \wedge r \Rightarrow q$
\end{quote}
which is logically equivalent to $r \Rightarrow (p \Rightarrow q)$.
Here $p$ and $q$ are taken from the method contract (\texttt{requires} $p$ 
\texttt{ensures} $q$) while $r$ is shown in the 
\enquote{Semantics} view of the method body (section \enquote{Transition Relation}).

This task shows the \emph{partial correctness} of the method, provided that
all preconditions and loop-related verification conditions (such as the
preservation of the loop invariant) are indeed correct (which is shown by
other tasks explained below). We therefore can detect in this
stage that the invariant of a loop in the body of the method
is too weak to show the partial correctness of the
method, even before the proof of the correctness of the invariant is actually
attempted. 

\paragraph{Termination} The goal of the proof that the \enquote{method body
terminates} is of form
\begin{quote}
$p \wedge \neg d \Rightarrow t$
\end{quote}
where $p$ represents the method's precondition, $d$ represents the
\texttt{diverges} condition in the method's contract (an optional condition
under which the method is allowed to run forever) and $t$ is the method body's
termination condition as depicted in the method's \enquote{Semantics} view 
(Section~\enquote{Termination Condition}). 

This task shows (together with task \enquote{postcondition}) the \emph{total
correctness} of the method, provided that the body of every loop in the method
terminates and that the associated termination term is well-formed and decreased
(which is shown by other tasks explained below). We can detect in this task that
a termination term has initially a wrong (negative integer) value or that some
method is called in a state in which the negation of its \texttt{diverges}
condition does not hold.

\paragraph{Preconditions} For every command with pre-state knowledge $k$ and
precondition $c$, a task is generated to show that the pre-condition is met, 
i.e.\ to prove a goal of form
\begin{quote}
  $k \Rightarrow c$
\end{quote}
where both $k$ and $c$ are displayed in the \emph{Semantics} view of the method
(sections \enquote{Pre-State Knowledge} and \enquote{Precondition}). We have
chosen this \enquote{top-down} generation of preconditions over the
\enquote{bottom-up} calculation of the calculus presented in
Section~\ref{theory} in order to foster a closer and more illustrative
relationship between a precondition and its statement respectively the
associated pre-state knowledge.

We can thus detect in this task that a command is executed in a state in which
the consequence of the execution may not be properly
described by the command's state relation.

\paragraph{Loops} For a while loop with an invariant $i$ and a body with state
relation $r$, there are four tasks generated (compare with the corresponding
rules in Section~\ref{theory}).

The first task is to verify the correctness of the invariant amounts 
to proving a formula of form 
\begin{quote}
  $i' \wedge e \wedge r \Rightarrow i''$
\end{quote}
Here $i'$ represents a variant of $i$ that expresses the relationship between
the initial state~$x$ of the loop and the state $y$ before the current loop
iteration, $e$ expresses the fact that the loop condition holds at state~$y$,
$r$ expresses the relation between the states~$y$ and~$z$ before and after the
current loop iteration, and $i''$ represents a variant of $i$ that expresses the
relationship between states~$x$ and~$z$:
\begin{quote}
\begin{picture}(0,0)%
\includegraphics{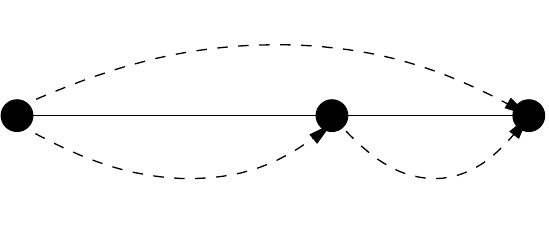}%
\end{picture}%
\setlength{\unitlength}{3315sp}%
\begingroup\makeatletter\ifx\SetFigFont\undefined%
\gdef\SetFigFont#1#2#3#4#5{%
  \reset@font\fontsize{#1}{#2pt}%
  \fontfamily{#3}\fontseries{#4}\fontshape{#5}%
  \selectfont}%
\fi\endgroup%
\begin{picture}(3121,1291)(1703,-1610)
\put(1801,-1321){\makebox(0,0)[b]{\smash{{\SetFigFont{10}{12.0}{\familydefault}{\mddefault}{\updefault}{\color[rgb]{0,0,0}$x$}%
}}}}
\put(3601,-1321){\makebox(0,0)[b]{\smash{{\SetFigFont{10}{12.0}{\familydefault}{\mddefault}{\updefault}{\color[rgb]{0,0,0}$y$}%
}}}}
\put(4726,-1321){\makebox(0,0)[b]{\smash{{\SetFigFont{10}{12.0}{\familydefault}{\mddefault}{\updefault}{\color[rgb]{0,0,0}$z$}%
}}}}
\put(3601,-826){\makebox(0,0)[b]{\smash{{\SetFigFont{10}{12.0}{\familydefault}{\mddefault}{\updefault}{\color[rgb]{0,0,0}$e$}%
}}}}
\put(3331,-466){\makebox(0,0)[b]{\smash{{\SetFigFont{10}{12.0}{\familydefault}{\mddefault}{\updefault}{\color[rgb]{0,0,0}$i''$}%
}}}}
\put(2791,-1546){\makebox(0,0)[b]{\smash{{\SetFigFont{10}{12.0}{\familydefault}{\mddefault}{\updefault}{\color[rgb]{0,0,0}$i'$}%
}}}}
\put(4231,-1546){\makebox(0,0)[b]{\smash{{\SetFigFont{10}{12.0}{\familydefault}{\mddefault}{\updefault}{\color[rgb]{0,0,0}$r$}%
}}}}
\end{picture}%

\end{quote}

The other three tasks are related to showing the termination of the loop.
The task to show that the loop \enquote{body terminates} is 
essentially to prove a goal
\begin{quote}
$i' \wedge e \Rightarrow b$
\end{quote}
where $i'$ and $e$ are as indicated  
above and $b$ is the termination condition derived from the loop body.
The task to show that the loop \enquote{measure is well-formed} is 
essentially to prove a goal
\begin{quote}
$i' \wedge e \wedge r \Rightarrow 0\leq t'$
\end{quote}
where $i'$, $e$, and $r$ are as indicated  
above and $t'$ represents the value of the termination term after the  
iteration of the loop. The task to show that the loop 
\enquote{measure is decreased} is essentially to prove a goal
\begin{quote}
$i' \wedge e \wedge r \Rightarrow t'<t$
\end{quote}
where $i'$, $e$, $r$, and $t'$ are indicated as above and $t$ represents
the value of the termination term before the iteration of the loop.

\paragraph{Specification Validation} An open question in every formal 
verification is whether the specification indeed expresses the informal
requirements that the programmer wants to impose on the program. A typical
beginner's error is that a specification 
(due to some error in the logical formulation, e.g., some wrong
logical connective) admits \emph{every} possible output
from an implementation. Given a specification with input variable $x$ and 
output variable $y$, precondition $p$ and postcondition $q$, 
the verification condition
\[
  \forall x: p(x) \Rightarrow \exists y: \neg q(x,y)
\]
rules such a \enquote{trivial} specification out. Likewise, the verification 
condition
\[
  \forall x: p(x) \Rightarrow \exists y: q(x,y)
\]
ensures that the specification allows \emph{some} output, i.e.\ that it is
actually \enquote{satisfiable} (implementable by a method). Furthermore, we may
show that for some concrete input $i$ some desired output $o$ is 
indeed legal by proving
\[
  p(i) \wedge q(i, o)
\]
respectively we may show that some undesired output $o'$ is illegal
\[
  p(i) \wedge \neg q(i, o')
\]
An extensive validation of every specification is recommended before
any of the previously described verification tasks is attempted.

The first two kinds of conditions are are generated by the RISC ProgramExplorer
for every method specification; $p$ and $q$ are derived from the method
contract and the roles of $x$ and $y$ are taken by variables that represent a
method body's pre- and post-state. The other two kinds of conditions will also
be generated in a future version of the software.

\section{General Workflow} 
\label{workflow}

The RISC ProgramExplorer has been designed to support the following 
workflow that leads in a systematic way from an informal problem statement
to a formal problem specification and subsequently to
a problem solution that is verified to be correct
with respect to the specification:

\begin{enumerate}

\item \textbf{Theory Development:} Considering the particular domain of the
problem at hand, we formalize in the RISC ProgramExplorer a corresponding
mathematical theory by defining or axiomatizing constants, functions, and
predicates that are suitable to express the concepts that are of interest in
that domain. We also formulate prospective theorems that may become useful as
knowledge about these concepts in the subsequent verification tasks. We may
prove these theorems immediately or delegate the proofs to a later stage.

\item \textbf{Method Specifications:} Considering the particular computational
problem in the domain, we describe the solution to the problem by a method
signature (method declaration with empty body) and a specification of
the method with the help of the concepts introduced in the
previously developed theory. We validate the specification by showing that
it is non-trivial, satisfiable, and holds for certain legal
input/output pairs (respectively does not hold for certain illegal pairs).

\item \textbf{Method Design:} We sketch a method solution by providing
a skeleton of the method body; the body may contain loop skeletons and
calls of new auxiliary methods that are specified by loop invariants, termination
terms, respectively method contracts. The loop respectively method bodies
can be implemented at a later stage (a loop body must however indicate by 
dummy assignments which variables are to be changed by an iteration).

\item \textbf{Semantic Analysis:} Based on the sketch of the method body
(and the specification of the method, the loops executed, and other methods
called), we investigate the semantic essence of the method. Do
the derived state relations represent the envisioned behavior? Are the
command's preconditions and termination conditions implied by the 
respective pre-state knowledge of the method? If some problem is
detected, we revise the method specifications, loop invariants, etc.

\item \textbf{Verification:} We attempt to verify the method's postcondition (partial
correctness). If the proof fails, it may be necessary to
revise the implementation, specifications, loop invariants. It may be also
necessary to extend the theories by introducing additional theorems that
provide knowledge that is required in the verification. 

Once the proof succeeds, we
attempt to verify the method's termination in the same manner (which
may lead to a revision of the termination terms).

Once these \enquote{major} verification tasks have been successfully completed,
we may turn to the \enquote{minor} tasks such as proving the statement 
preconditions and the loop-related tasks (correctness of invariants and
termination terms). If failures are detected, again the implementation,
specifications, loop invariants, termination terms, or theories may need
to be revised.

\item \textbf{Refinement:} We refine the still open bodies of loops and
auxiliary methods; we analyze, and verify them, as described above.

\item \textbf{Theorems:} We prove the still open theorems that were used in the 
verification process (respectively were newly
introduced in the course of this process).

\end{enumerate}

In this methodology (which is iteratively performed if problems/errors
are detected at a certain stage), the actual program verification is a core step, but not the
only one. Most important, the semantic analysis \emph{precedes} the
verification; programmers should thus get insight into the program methods and
their specification \emph{before} they dive into the depths of their formal
verification. The verification itself is also organized in a layered
fashion where e.g.\ a proof of partial correctness is performed before the
correctness of the loop invariants or of the contracts of auxiliary methods has been
established; the suitability of the invariants respectively contracts for the
task at hand can thus be tested at an early stage. It should be also noted that
the process integrates the principle of \emph{refinement} (respectively
\enquote{correct by construction}) by deferring the implementation of loop
respectively method bodies to later stages of the development.

\section{Conclusions}
\label{conclusions}

The approach to program reasoning presented in this paper and its implementation
in the RISC ProgramExplorer were motivated by a personally experienced lack of
transparency in existing tools which made it hard for the
author to get deeper insight into a program from the automatically generated
verification conditions and the (failed) attempts to prove them. Our goal was to
make not only the derivation transparent but to base this derivation on a
semantic model that can be presented to the user and is open for further
investigation (prior to any actual verification attempt). For this purpose, we
have constructed a denotational semantics of a program that is based on the
model of a program as a state relation; this relation can be described by a
classical logical formula and can serve in a quite direct way as the basis of a
verification of the program. Since the relations are derived from the
specifications of the methods called and the loops executed (not from the bodies
of the methods respectively loops), the approach is inherently modular and gives
rise to a step-wise refinement of program specifications to implementations.

In a certain sense, the presented approach can be considered as the translation
of an imperative model of programming to a declarative one. In the imperative
model, the focus is on a sequence of variable updates that gradually transforms a store
from a given initial situation to a desired terminal situation; in the
declarative model, the focus is on the relationship between the
given and the desired situation. In our experience, most students think of
computational problems mainly in an imperative view
and are insecure with the declarative view (which is however the basis of formal
reasoning), i.e., they tend to think in terms of \enquote{how} rather than in
terms of \enquote{what}. The difference is predominant in the typical programs
that express repetitive computations in the form of loops (if a program is
written in a functional style where recursion is applied for this purpose, the
difference between both views becomes blurred). The presented translation is
designed to help people that are trained in the imperative view to become
also proficient with the declarative one.

Consequently the RISC ProgramExplorer was developed to provide a close integration between
programs, theories, specifications, and semantic models in order to emphasize
the co-development of the declarative view of a problem (its specification) and
the operational view (its implementation) and to exhibit the relationship
between them. To support actual verifications, the RISC ProofNavigator was
designed as a compromise between a certain level of automation (which is
necessary to perform proofs successfully) and a comfortable user interaction
(which is necessary to direct the prover into the right direction and, in
particular, to learn from failed proof attempts). A particular challenge was the
appropriate simplification of the automatically derived transition formulas to a
form that a human would consider as the most \enquote{natural} one. First
experiments seem to indicate that by appropriate simplification of these
formulas (flattening quantifier structures, eliminating variables, etc.), also
the consequent verifications become technically simpler; we consider this as an
interesting problem that needs further research.

The RISC ProofNavigator has been applied since 2005 in a regular course on
\enquote{Formal Methods in Software Development} for the proof of (manually
derived) verification conditions. A detailed presentation of our experience is
beyond the scope of this paper (see also~\cite{Schreiner2008}; in a nutshell, 
we found that most students
become able, after a comparatively short repetition of the basics of logic and
proving (some prior background in logic is assumed) and a corresponding
introduction to the system and its user interface, to perform verifications of
correct programs with given correct specifications and program annotations (loop
invariants and termination terms). However, many tend to have big problems if
programs, specifications, and/or annotations contain errors (the less bright/motivated
ones then give quickly up or perform seemingly random proving commands). 

The RISC ProgramExplorer and the methodology it supports are being tested for
the first time in the current iteration of a course on \enquote{Formal Methods
in Software Development} that has started at the Johannes Kepler University Linz
in October 2011; this course is mandatory for students of the master programmes
\enquote{Software Engineering} and \enquote{Computer Mathematics}. In previous
iterations, we have experienced that many students did not really get deeper
insight into e.g.\ the expressiveness of loop invariants and their role and
suitability with respect to proving the partial correctness of a method. We hope
that by the new tool and the corresponding methodology, this insight will be
substantially deepened. A critical point will certainly be the ability to deal
with the various views on a program and to relate them to each other. Our experience will
show to which extent our idea will be successful and also give feedback for the
further evolution of our software and for the development of an accompanying
didactic approach. 

\providecommand{\urlalt}[2]{\href{#1}{#2}} 
\providecommand{\doi}[1]{doi:\urlalt{http://dx.doi.org/#1}{#1}}

%\nocite{*}
\bibliographystyle{eptcs}
\bibliography{bibliography}
\end{document}